\documentstyle[12pt]{report}
\begin{document}
\baselineskip=0.8cm
\ \ \

\begin{center}
{\bf DEEP IR IMAGING OF GLOBULAR CLUSTERS. III. M13}
\end{center}

\begin{center}
T. J. Davidge$^a$, Gemini Canada Project Office,

Dominion Astrophysical Observatory, 5071 W. Saanich Road,

Victoria, B.C. CANADA V8X 4M6$^b$

and

Department of Geophysics \& Astronomy, University of British Columbia,

Vancouver, B.C., CANADA V6T 1Z4

email: davidge@dao.nrc.ca
\end{center}

\begin{center}
W. E. Harris$^a$, Department of Physics \& Astronomy,

McMaster University, Hamilton, Ontario, CANADA L8S 4M1

email: harris@physun.physics.mcmaster.ca
\end{center}

\noindent{To appear in The Astrophysical Journal}

\vspace{1.5cm}
\noindent{$^a$} Visiting Astronomer, Canada-France-Hawaii Telescope, which is
operated by the National Research Council of Canada, the Centre National de la
Recherche Scientifique de France, and the University of Hawaii.

\vspace{1.5cm}
\noindent{$^b$} Mailing address

\pagebreak[4]
\begin{center}
{\bf ABSTRACT}
\end{center}

	We have used the CFHT REDEYE infrared camera to obtain deep $J$ and
$K'$ images of four fields in the globular cluster M13. The composite
$(K, J-K)$ color-magnitude diagram (CMD) extends from the upper red giant
branch (RGB) to 2 magnitudes below the main sequence turn-off (MSTO). Selected
[Fe/H] $\sim -1.6$ isochrones from Bergbusch \& VandenBerg (1992, ApJS, 81,
163) and Straniero \& Chieffi (1991, ApJS, 76, 525) are transformed onto the
near-infrared observational plane, and these suggest an age for M13 in the
range $14 - 16$ Gyr. We emphasize that any effort to estimate ages should be
considered as tentative given uncertainties in the input physics;
however, these uncertainties notwithstanding, comparisons between the
near-infrared isochrones indicate that age {\it differences} of order $\pm 2$
Gyr should be detectable among metal-poor clusters on the $(K, J-K)$ plane.
Building on this result, we find that the difference in $J-K$ color
between the MSTO and the base of the RGB in M13 is the same as in M4, and
conclude that these clusters have similar ages. This conclusion is verified by
comparing (1) the $K$ brightnesses of the MSTO, and (2) the morphologies of
optical CMD's. Finally, we investigate the mass function of main sequence
stars in M13 with distances between 3 and 6 core radii from the cluster
center. The mass functions in the interval 0.55 and 0.8 M$_{\odot}$ are
relatively flat, independent of radius. Optical studies at larger
radii have found non-zero mass function indices in this same mass interval,
and we attribute this radial variation in mass function morphology to
dynamical evolution.

\vspace{0.3cm}
\noindent{\it Subject Headings:} globular clusters: individual (M13) $-$
infrared: stars $-$ stars: evolution $-$ stars: luminosity function, mass
function

\pagebreak[4]
\begin{center}
I. INTRODUCTION
\end{center}

	Because it is among the closest globular clusters, and is located at a
moderately high Galactic latitude, M13 has been the target of several
photometric studies. Early photographic and
photoelectric investigations by Kadla, Zdarsky, \& Spasova (1976), Sandage
(1970), Baum et al. (1959), Savedoff (1956), Arp (1955), Arp \& Johnson (1955),
and Baum (1954) defined the cluster color-magnitude diagram (CMD) from the
upper red giant branch (RGB) to well below the main sequence turn-off (MSTO).
More recent CCD studies by Richer \& Fahlman (1986) and Lupton
\& Gunn (1986) have traced the cluster sequence to $V \sim 24$,
which corresponds to stars with masses near 0.2M$_{\odot}$.

	Although there is a wealth of optical photometric data, only a
small number of M13 stars have been observed
at wavelengths longward of 1$\mu$m. Cohen, Frogel, \& Persson (1978)
obtained aperture measurements of 14 bright giants, and these data were used
as part of a larger effort to calibrate giant branch properties as a function
of metallicity (Frogel, Cohen, \& Persson 1983).
More recently, Buckley \& Longmore (1992) obtained $K$ photometry of a small
number of main sequence stars and, after combining these data with optical
measurements, estimated basic properties for M13 such as temperature of
the MSTO and distance. The absence of a larger set of
near-infrared measurements is unfortunate since,
by virtue of the extensive optical photometric database, coupled with the
broad agreement which exists between the various
spectroscopic metallicity estimates (Section 4),
M13 is an ideal calibrator for near-infrared isochrones. The issue of isochrone
calibration longward of 1$\mu$m is important, as efforts to
transform theoretical isochrones onto the near-infrared
observational plane using bolometric corrections and color calibrations
predicted by model atmospheres have revealed surface gravity and
temperature-dependent errors, which distort the transformed models (e.g.
Davidge \& Simons 1994a,b $-$ hereafter referred to as Papers 1 and 2;
Ferraro et al. 1994).

	Aside from studies of CMD morphology, deep infrared photometry can also
be used to derive mass functions independent of those computed at optical
wavelengths. The mass function of M13 is of particular interest since studies
by Richer \& Fahlman (1986), Lupton \& Gunn (1986), Drukier et al. (1988) and
Richer et al. (1990), suggest that at low masses the M13 mass function may be
steeper than in the solar neighborhood. Consequently, M13 should
play an important role in defining correlations between mass function index
and quantities such as metallicity and distance from the Galactic Centre (e.g.
Djorgovski, Piotto, \& Capaccioli 1993). The study of the M13 mass function
is also significant since star counts suggest that the mass function of
halo field stars may follow that of clusters (Reid \& Majewski 1993).

	M13 should also prove to be an excellent comparison object for
metal-poor clusters in the Galactic Bulge which, according to some models of
spiral galaxy formation (e.g. Larson 1990), may be the oldest in the Galaxy.
Indeed, VandenBerg, Bolte, \& Stetson (1990)
included M13 in their sample of [Fe/H] $\sim -1.6$ halo clusters for
relative age determinations and, based on the Richer \& Fahlman (1986)
photometry, suggested that M13 may be the oldest cluster in this group,
although the lack of a large database of sub-giant branch (SGB) and lower RGB
stars with CCD photometry prevented definitive conclusions from being drawn.
Near-infrared photometry is required for studies of Bulge clusters, as these
objects are often heavily reddened, and spatial variation
in the distribution of obscuring material smears optical
CMD's. Fortunately, these effects are greatly
reduced at wavelengths longward of 1$\mu$m, and $(K, J-K)$ CMD's should prove
useful for probing age differences. In future papers of this series, we will
compare our M13 CMD with those of Bulge clusters.

	In the current paper, we discuss $J$ and $K'$ observations of four
fields in M13. The main goals of this study are (1) to define the $(K, J-K)$
CMD of M13 over a wide range of brightnesses, (2) to compare the CMD with
evolutionary models, and (3) to investigate the cluster mass function.
The observations and reductions are described in Section 2,
while details of the photometry and the resulting
$(K, J-K)$ CMD are discussed in Section 3. In Section 4 we compare the CMD with
theoretical isochrones, which have been translated onto the near-infrared
observational plane using a technique that relies on color relations defined by
metal-poor sub-dwarfs and giants, combined with synthetic photometry predicted
from model atmospheres. We investigate the mass function of M13 in Section 5,
and a brief summary follows in Section 6.

\pagebreak[4]
\begin{center}
II. OBSERVATIONS AND REDUCTIONS
\end{center}

	The data were recorded during the nights of UT August 28, 29, and 31
1993 using the Canada-France-Hawaii Telescope (CFHT) REDEYE camera, which was
mounted at the f/8 focus of the telescope. The camera contains a Rockwell
256 x 256 Hg:Cd:Te array with 40$\mu$m square pixels;
a detailed description of the instrument has been given by Simons et al.
(1993).
The data were recorded through the narrow-field optics, so that the
image scale is 0.2 arcsec/pixel and each exposure samples a field
51 arcsec on a side. Because of the high background
levels inherent to broad-band near-infrared observations, multiple exposures
were recorded of each field, and the telescope was offset a few arcsec between
these (ie. `dithering') to facilitate the identification and suppression of
bad pixels. Consequently, the final co-added images cover regions slightly less
than 51 arcsec on a side.

	Four cluster fields were observed through a Caltech-CTIO $J$
and a $K'$ (Wainscoat \& Cowie 1992) filter. The co-ordinates of each field,
as well as the integration times and image quality in the final frames
are summarized in Table 1. A number of standard stars from the list compiled
by Casali \& Hawarden (1992) were also observed, and these data
will be discussed further in Section 3.

	A problem developed with the array controller while the Field
3 data were being recorded, such that the read noise in one quadrant became
very high. This had a significant impact on photometry of
faint sources, so this portion of each Field 3 image
was excluded from the final analysis. However, the
photometric performance of the other quadrants was unaffected.

	The data were reduced using procedures similar to those described in
Papers 1 and 2. A median dark frame was subtracted from each
image, and the results were divided by normalized sky flats, constructed on a
nightly basis from all deep integrations recorded on that night. The
flat-fielded images were sky subtracted, aligned, median combined,
and then trimmed to the area common to all exposures. The final $J$ images
are shown in Figure 1.

\begin{center}
III. PHOTOMETRIC MEASUREMENTS AND THE $K, J-K$ DIAGRAM
\end{center}

\noindent{\it 3.1 Photometric Measurements}

	The brightnesses of standard stars were measured using the aperture
photometry routine PHOT, which is part of the
DAOPHOT (Stetson 1987) photometry package. Aperture sizes
were determined by inspecting radial intensity profiles of the various
standard stars. Linear transformations of the form:

\hspace*{4.0cm}$M = m + \psi_M C + \mu_M X + \zeta_M$

\noindent{were} then fit to the measurements using the method of least squares.
Here, $M$ and $m$ are the standard and instrumental
magnitudes, $C$ is the instrumental color,
$X$ is the airmass, $\psi_M$ and $\zeta_M$ are the transformation coefficients,
and $\mu_M$ is the extinction coefficient. The instrumental $K'$ measurements
were transformed into $K$ values.

	The extinction coefficients in $J$ and $K$ are relatively small, and
require a large number of standard stars to be reliably determined. A total of
20 standard star observations were made during the course of
the five night observing run. Because of this modest sample size,
it was decided to adopt the Mauna Kea $J$ and $K$ extinction coefficients
computed by Guarnieri, Dixon, \& Longmore (1991).
After fixing the extinction coefficients at these values, a least squares fit
to the tranformation relations revealed $(j - k')$ color coefficients of
$0.015 \pm 0.037$ and $0.020 \pm 0.056$ for $J$ and $K$, respectively. The
uncertainties in the derived zeropoints are $\sim 0.01$ magnitudes.
A list of the standards which were observed, and the residuals from the
fitted transformation relations, in the sense standard value minus computed
value, is given in Table 2.

	Photometry of cluster stars was performed with the PSF-fitting routine
ALLSTAR (Stetson \& Harris 1988). Artificial star experiments were run to
determine completeness fractions and assess systematic errors in the
measured brightnesses for each field. Typically $30 - 50$ artificial stars
per 0.5 magnitude interval were added to the various frames,
and this was done over the course of several runs to prevent
artificially raising the degree of crowding. The completeness curves for Fields
1 and 3, which have the shortest and longest integration times, are compared in
Figure 2.

\vspace{0.5cm}
\noindent{\it 3.2 The $(K, J-K)$ CMD of M13}

	The $(K, J-K)$ diagrams for all four fields are shown in Figure 3.
Only those stars with uncertainties in the computed brightnesses due to
errors in the PSF-fit, as calculated by
ALLSTAR, $\leq 0.07$ magnitudes in each filter are plotted.
Initial comparisons revealed that the loci of the Field 1 and 4 CMDs agreed to
within 0.02 magnitude; however, the CMDs for Fields 2 and 3 were offset by
$\sim 0.07$ magnitudes in $(J-K)$ from the Field 1 and 4 data, in the sense
that the Field 2 data were too red, whereas the Field 3 data were too blue.
We are unable to offer an unambiguous explanation for these offsets, although
the standard star data provide a clue to the likely cause, which we now
discuss.
The data for Fields 1, 2 and 3 were recorded largely on the night of August 28.
Based on the data in Table 2, the standard deviation in $\Delta_{JK} = \Delta_J
- \Delta_K$ among the standard stars observed on this night is $\pm 0.07$
magnitudes, which is remarkably (and probably fortuitously!) similar to the
observed field-to-field scatter in the color
calibration. For comparison, the Field 4 data were
recorded exclusively on the night of August 31, and the standard
deviation in $\Delta_{JK}$ for the standard stars observed on this night is
only 0.03 magnitudes. This comparison of the standard star data suggests that
modest changes in sky transparancy may be responsible for the offsets in color
calibration. To compensate for these, the Field 2 and 3
data were shifted to match those of Fields 1 and 4, and it
is the corrected measurements for Fields 2 and 3 which are shown in Figure 3.

	An independent check of our color calibration is highly desireable
given the field-to-field variation noted in the preceding paragraph. To
facilitate such a check, normal points were created from the data in Figure 3
by sorting the observations into bins of width $\pm 0.25$ magnitude in $K$, and
then calculating the mean of the $J-K$ distribution in each bin after applying
sequencial $2 - \sigma$ rejections. The Field 1 data were used to define the
cluster sequence above the MSTO, while the Field 2, 3, and 4 data were used for
the SGB and main sequence. The results are listed in Table 3. Also shown in
Table 3 are the estimated uncertainties in the normal
points, which are the standard errors of the mean computed
from the $J-K$ distribution in each brightness bin following the sequential
rejection iterations described above. We emphasize that these uncertainties are
internal, and that additional systematic uncertainties in the calibration
amounting to $\sim 0.01 - 0.02$ magnitudes are also present.

	In Figure 4 the normal points are compared with
the aperture measurements made by Cohen et al. (1978)
and the ridgeline predicted from the measurements
made by Buckley \& Longmore (1992). Buckley \& Longmore obtained $K$, but not
$J$, measurements, and computed $V - K$ colors from existing optical
photometry. In order to create measurements which could be compared with our
data, $J-K$ colors were calculated using the $(J-K, V-K)$ calibration
for metal-poor sub-dwarfs derived in Section 4.
It is evident from Figure 4 that both the Cohen et al. and the Buckley \&
Longmore data are in good agreement with our observations, adding
confidence to our color calibration.

	How does the near-infrared CMD of M13 compare with those of other
metal-poor clusters? The ridgeline of the [Fe/H] $\sim -1.3$ (Zinn \& West
1984) cluster M4, taken from Paper 1 and shifted to match the color and
brightness of the M13 sequence in the vicinity of the MSTO, is compared with
our normal points in Figure 4. Considering the uncertainties in the M13
normal points, there is reasonable agreement between the two
sequences in the vicinity of the MSTO, and the $(J-K)$ color
difference between the MSTO and the base of the RGB in the two clusters is
identical, suggesting similar ages. If M4 and M13 are coeval, then their MSTO
brightnesses should also be similar, although a comparison of this nature is
complicated by uncertainies in reddening and differences in metallicity. The
near-infrared CMD of M4 presented in Paper 1 indicates that K$_{MSTO} \sim
14.25$, while in M13 K$_{MSTO} \sim 17.0$, with estimated uncertainties of
$\pm 0.10$ magnitudes. Based on data given in Table 4 of Armandroff (1989),
$E(B-V) = 0.40$ and V$_{HB} = 13.35$ for M4, while $E(B-V) = 0.02$ and V$_{HB}
= 14.95$ for M13. Adopting the RR Lyrae brightness calibration predicted by
Lee, Demarque, \& Zinn (1990), and the reddening curve tabulated by Rieke \&
Lebofsky (1985), such that A$_{K} = 0.112$A$_{V}$ and $E(J-K) = 0.52 E(B-V)$,
then M$_{K}^{MSTO} = 2.60$ in M4 and 2.65 in M13. Theoretical near-infrared
isochrones constructed using the procedures described in Section 4 predict that
$\Delta$M$_{K}^{MSTO}/\Delta$t = 0.1 magnitudes/Gyr, and
$\Delta$M$_{K}^{MSTO}/\Delta$[Fe/H] = 0.1 magnitudes/[Fe/H] dex.
Consequently, the near-infrared brightnesses of the MSTO
are also consistent with similar ages for M4 and M13.

	The hypothesis that M4 and M13 have similar ages can be tested further
using optical data. To this end, we compared the M4 normal points
derived by Alcaino, Liller, \& Alvarado (1988) with those for M13
from Richer \& Fahlman (1986), and the result is shown in Figure 5 where, as
in Figure 4, the M4 data have been shifted to match the color and brightness of
the M13 MSTO. Although the lower RGB is poorly defined in both clusters,
there appears to be general agreement between the M4 and M13 sequences on
the $(V, B-V)$ plane. Clearly, additional optical observations covering the
SGB and lower RGB will be required to confirm this result.
Nevertheless, given that age differences of $\pm 2$ Gyr should be detectable
between metal-poor clusters in the $K, J-K$ plane (Section 4), we tentatively
conclude that M13 and M4 have similar ages. We note that a tacit assumption
behind this conclusion is that the abundances of key atomic species, such as
oxygen, do not vary among clusters with similar [Fe/H], since differences in
chemical composition can mimic age effects on the CMD (VandenBerg \& Stetson
1991).

\begin{center}
IV. COMPARISON WITH THEORETICAL MODELS
\end{center}

	Before comparing the observations with theoretical sequences, we
briefly review metallicity estimates for M13, with emphasis on spectroscopic
studies. The main purpose of this exercise is to determine the mean metallicity
of the cluster, so that this parameter can be fixed for comparisons
with isochrones. Cohen (1978), Leep, Wallerstein, \& Oke (1986),
Wallerstein, Leep, \& Oke (1987), and Lehnert, Bell, \& Cohen (1991)
obtained spectra of bright M13 giants, and concluded that [Fe/H]
lies between $-1.5$ and $-1.6$, with a mild preference for values at
the lower end of this range. The sole discrepant spectroscopic study
is that by Pilachowski, Wallerstein, \& Leep (1980), who found [Fe/H]
$\sim -1.42$ based on 5 giants. This result aside,
the spectroscopic data are consistent with
the mean metallicity computed by Zinn \& West (1984), who used a variety of
indicators to find [Fe/H] $= -1.65 \pm 0.06$. Consequently, in the following
we will compare our observations with models having [Fe/H] $\sim -1.6$.

	Although the photometric scatter in the optical CMD of M13 indicates
that the star-to-star metallicity dispersion must be small
(Folgheraiter, Penny \& Griffiths 1993), there is evidence for a range of CNO
abundances, at least among the most evolved stars. This is potentially
important when comparing observations with evolutionary models, as the
abundances of species such as oxygen can influence
isochrone morphology (e.g. VandenBerg \& Stetson 1991). Suntzeff (1981)
and Langer, Suntzeff, \& Kraft (1992) found that giants in M13
show a range of CN strengths at a given $B-V$, with the majority
being CN-strong. Leep et al. (1986), Brown, Wallerstein, \& Oke (1991), and
Kraft et al. (1992, 1993) find a range of oxygen abundances,
although most of the M13 giants they studied appear to be oxygen
{\it deficient} (ie. [O/Fe] $\leq 0$). This is very different from what is
seen among field giants and dwarfs (e.g. Abia \& Rebolo 1989;
Barbuy \& Erdelyi-Mendes 1989); nevertheless, given the evolved nature
of these stars, it is possible that the M13 studies may not be
measuring primordial abundances, but rather may be seeing the results
of mixing. In the following, we will consider models which make different
assumptions concerning [O/Fe].

	In Papers 1 and 2 it was found that isochrones
transformed onto the near-infrared observational plane using relations derived
from model atmospheres contained luminosity-dependent calibration errors,
which are largest in the vicinity of the MSTO. Bell (1992) was the first to
detect these, and suggested that uncertainties in the H$^-$ free-free
opacity may be partly responsible. Nevertheless, once
corrected for calibration errors, certain sections of the transformed
isochrones, such as the RGB, are able to closely reproduce the observations.
In Paper 2 an effort was made to avoid calibration problems by constructing
empirical color relations, which were in turn used to derive M$_K$ and $J-K$
from the tabulated values of M$_V$ and $B-V$.
This procedure assumes that the optical isochrones do
not contain large calibration errors, an assumption which appears to be
valid (VandenBerg 1992). In fact, the near-infrared isochrones created in this
manner provide a reasonable match to the observations, although
a significant shortcoming is that there
are some parts of the CMD, such as the base of the RGB, where
the paucity of nearby calibrating stars makes it difficult
to derive optical-infrared color relations. In the current paper, we
describe a hybrid technique, which builds on the strengths of the
theoretical and empirical calibration methods, to
transform selected isochrones onto the near-infrared observational plane.

	The first step is to assemble empirical color relations
appropriate for [Fe/H] $\sim -1.6$. To this end, optical-infrared
color relations were derived for very low mass stars, bright sub-dwarfs, and
bright giants. The relations between $B-V$,
$J-K$ and $V-K$ for halo sub-dwarfs from Table 6 of Leggett (1992)
were adopted for stars near the bottom of the main sequence,
while color relations for brighter sub-dwarfs were derived from
data tabulated by Laird, Carney, \& Latham (1988) for stars with [Fe/H] between
$-1.5$ and $-2.0$. Color relations for bright giants were derived from the
photometry tabulated by Cohen et al. (1978), Cohen \& Frogel (1982), and
Frogel, Persson, \& Cohen (1983) for the clusters NGC 288, NGC 1904, NGC 2298,
NGC 4372, NGC 4833, NGC 5024, NGC 5272, NGC 5286, NGC 5904, NGC 6254, NGC 6752,
and NGC 7006 which, according to Zinn \& West (1984), span a range of
metallicities
similar to that used for the bright sub-dwarfs. The data used to derive the
bright sub-dwarf and giant color relations are summarized in
Figures 6 ($B-V$ vs $J-K$) and 7 ($B-V$ vs $V-K$), while
the adopted color relations are listed in Tables 4 and 5.

	The color relations do not cover the SGB and lower-RGB. These
gaps were filled by extrapolating the empirical color relations
using values of $\Delta(J-K)/\Delta(B-V)$ and $\Delta(V-K)/\Delta(B-V)$
derived from the synthetic photometry tabulated by Bell \& Gustafsson (1989)
and Bell (1992). We emphasize that although the synthetic photometry
contains significant calibration errors, these data are nevertheless able to
reproduce the {\it shape} of the RGB (e.g. Paper 1), so they should
provide reliable guides for extrapolation.

	The procedure described above was used to transform selected [Fe/H]
$\sim -1.6$ isochrones from Bergbusch \& VandenBerg
(1992), who assume non-solar oxygen abundances, and Straniero
\& Chieffi (1991), who assume [O/Fe] = 0, onto the near-infrared observational
plane. The 12 and 16 Gyr Bergbusch \& VandenBerg isochrones and 16 and 20 Gyr
Straniero \& Chieffi isochrones are compared in Figure 8. Age differences
are most evident among the Bergbusch \& Vandenberg (1992) isochrones, as these
extend to the youngest ages. The Bergbusch
\& VandenBerg isochrones predict a change in
the $J-K$ color of the MSTO of $\sim 0.05$ magnitudes between 12 and 16 Gyr,
while the brightness of the SGB changes by $\sim 0.6$ magnitudes.

	The theoretical sequences are compared with the M13 normal points in
Figures 9 and 10, while the Straniero \& Chieffi (1991) sequences are
compared with actual data from Fields 2 and 3 in
Figure 11. The placement of the models on these diagrams
assumes $E(J-K)=0.01$ and $\mu_0 = 14.29$, values which result if (1) $E(B-V) =
0.02$ and V$_{HB} = 14.95$ (Armandroff 1989); (2) M$_V^{HB} = 0.6$; and
(3) $E(J-K) = 0.52 E(B-V)$ (e.g. Rieke \& Lebofsky 1985). Neither set of models
provides an entirely satisfactory fit to the RGB;
nevertheless, in both cases the isochrones are able to match the main sequence
fainter than $K \sim 18$. The exact age infered for M13 depends on the models
that are used, although it appears that models with ages in the range
$14 - 16$ Gyr should provide the best match to the color of the MSTO,
and should be able to reproduce the brightness difference
between the SGB and MS at a fixed $J-K$. However, we emphasize that
shortcomings in the input physics of stellar structure models make the
determination of an absolute age uncertain. These uncertainties
notwithstanding, it is evident from Figures 8, 9, 10, and 11 that the $(K,
J-K)$ diagram could be used to detect age {\it differences} on the order of
$\pm 2$ Gyr among metal-poor clusters.

\begin{center}
V. LUMINOSITY AND MASS FUNCTIONS
\end{center}

	Lupton \& Gunn (1986), Drukier et al. (1988), and Richer et al. (1990)
investigated the mass function of main sequence stars in the outer regions
of M13. These studies found that (1) the number
of stars per mass interval increases with decreasing mass over all
parts of the main sequence; (2) there are discontinuities in the mass function
near $0.4 - 0.5$M$_{\odot}$, such that a single power law cannot fit the entire
range of masses; and (3) at low masses the mass function is steeper than in
the solar neighborhood.

	The completeness-corrected $K$ luminosity functions of main sequence
stars in Fields 2, 3, and 4 are shown in Figure 12, where $N_K$ is the number
of stars per 0.5 magnitude interval per square arcmin. The faint limit of the
luminosity functions was set at $K \sim 19$, the point at which a coherent
main sequence can no longer be traced in Figure 3. The luminosity functions
for Fields 2 and 4 are largely parallel; however, the Field 3 luminosity
function appears to be steeper than the others.

	A mass-luminosity relation derived from the near-infrared
16 Gyr Bergbusch \& VandenBerg (1992) isochrone generated in Section 4
was applied to the $K$ luminosity functions to create mass functions, and the
results are shown in Figure 13, where $N_M$ refers to the number of stars per
square arcmin per solar mass. The mass functions for Fields 2 and 4 appear to
be very flat down to log(M) $\sim -0.2$ (ie. $\sim 0.6$M$_{\odot}$). Although
the Field 3 mass function appears to be tilted with respect to others, this
finding rests largely on the most massive bins, where the uncertainties are
large due to small number statistics.

	In order to quantify the slopes of the mass functions, power-laws of
the form $N dM \hspace*{0.2cm} \alpha \hspace*{0.2cm} M^{x}dM$ were fit to the
curves in Figure 13. The entire 0.60 to 0.78M$_{\odot}$ range was included in
these fits, and the results are listed in Table 6. The results in this Table
verify the visual appearance of Figure 13, in that the mass function
exponents of Fields 2 and 4 are not significantly different from zero. For
comparison, the mass functions published by Lupton \& Gunn (1986), Drukier et
al. (1988), and Richer et al. (1990) at larger radii suggest that the mass
function exponent in this same mass interval is significantly different
from zero. This radial variation in mass function morphology is qualitatively
consistent with dynamical evolution. Pryor, Smith, \& McClure (1986) modelled
mass segregation in globular clusters and, if the concentration index for
M13 is C=1.5, as claimed by these authors, the mass function exponent
between 3 and 6 core radii should change by $\sim 0.5 - 1.0$ if the primordial
mass spectrum was like that in the solar neighborhood. It is apparent from
the uncertainties listed in Table 6 that a change in mass function exponent
of this size would go undetected with the current dataset.

\begin{center}
VI. SUMMARY
\end{center}

	The main conclusions of this paper are as follows:

\noindent{1)} The $(K, J-K)$ CMD's of M13 and M4 are very similar
in the vicinity of the MSTO and lower RGB, suggesting that
these clusters have comparable ages. This result is consistent with existing
CCD data, although there is a need to better define the SGB's of these
clusters at optical wavelengths.

\vspace{0.5cm}
\noindent{2)} Selected [Fe/H] $\sim -1.6$ isochrones from Bergbusch \&
VandenBerg (1992) and Straniero \& Chieffi (1991) have been transformed onto
the near-infrared observational plane using a technique which relies on
empirical color relations derived from metal-poor stars
and synthetic photometry derived from model atmospheres. A comparison of
near-infrared isochrones suggests that age differences of order $\pm 2$
Gyr can be detected among metal-poor clusters on the $(K, J-K)$ plane.

\vspace{0.5cm}
\noindent{3)} A comparison with [Fe/H] $\sim -1/6$ near-infrared isochrones,
which make different assumptions for [O/Fe], suggests that M13 has an age in
the range $14 - 16$ Gyr. This result should be viewed as tentative given
uncertainties in our knowledge of stellar evolution (e.g. VandenBerg 1992).

\vspace{0.5cm}
\noindent{4)} The mass function of stars in the mass interval $0.6 -
0.8$M$_{\odot}$ is flat near 4 core radii, whereas optical studies at larger
radii covering the same mass interval have found a mass function exponent
different from zero. Our results are qualitatively
consistent with M13 having undergone dynamical evolution.

\vspace{2.0cm}
	The authors gratefully acknowledge financial support from the Natural
Sciences and Engineering Research Council of Canada (NSERC) and the
National Research Council of Canada (NRC). An anonymous referee also provided
comments which improved the meanuscript.

\pagebreak[4]
\begin{center}
TABLE 1

SUMMARY OF OBSERVATIONS
\end{center}

\begin{center}
\begin{tabular}{ccccrr}
\hline\hline
 & & & & & \\
Field & R.A. & Dec. & R$_{center}^a$ & Exposures & FWHM \\
 & & & & & \\
 & (1950.0) & (1950.0) & (Core radii) & (sec) & (arcsec) \\
 & & & & & \\
\hline
 & & & & & \\
1 & 16:39:54.4 & +36:33:31.5 & 0.4 & 3 x 10 $(J)$ & 0.5 \\
 & & & & & \\
 & & & & 3 x 10 $(K')$ & 0.5 \\
 & & & & & \\
 & & & & & \\
2 & 16:40:01.1 & +36:36:06.1 & 3.9 & 3 x 90 $(J)$ & 0.5 \\
 & & & & & \\
 & & & & 3 x 90 $(K')$ & 0.5 \\
 & & & & & \\
 & & & & & \\
3 & 16:39:32.1 & +36:31:01.0 & 5.9 & 20 x 180 $(J)$ & 0.6 \\
 & & & & & \\
 & & & & 20 x 180 $(K')$ & 0.6 \\
 & & & & & \\
 & & & & & \\
4 & 16:39:40.2 & +36:33:59.5 & 3.4 & 50 x 60 $(J)$ & 0.6 \\
 & & & & & \\
 & & & & 45 x 60 $(K')$ & 0.6 \\
 & & & & & \\
\hline
\end{tabular}
\end{center}

\noindent$^a$ Distance from cluster center, assuming core radius = 49.9
arcsec (Webbink 1985).

\pagebreak[4]
\begin{center}
TABLE 2

STANDARD STAR RESIDUALS
\end{center}

\begin{center}
\begin{tabular}{lrrrrr}
\hline\hline
 & & & & & \\
Date & Star$^a$ & $J$ & $K'$ & $\Delta_J$ & $\Delta_K$ \\
 & & & & & \\
\hline
 & & & & & \\
Aug 28 & 23 & 12.997 & 12.374 & $+0.030$ & $-0.024$ \\
 & & & & & \\
 & 27 & 13.494 & 13.123 & $+0.017$ & $-0.054$ \\
 & & & & & \\
 & 34 & 12.819 & 12.989 & $-0.064$ & $-0.067$ \\
 & & & & & \\
 & 30 & 11.923 & 12.015 & $-0.074$ & $+0.062$ \\
 & & & & & \\
Aug 29 & 24 & 10.904 & 10.753 & $-0.008$ & $-0.076$ \\
 & & & & & \\
 & 25 & 10.231 & 9.756 & $+0.045$ & $+0.085$ \\
 & & & & & \\
 & 29 & 13.175 & 13.346 & $-0.018$ & $+0.036$ \\
 & & & & & \\
 & 2 & 10.713 & 10.466 & $+0.008$ & $-0.021$ \\
 & & & & & \\
 & 4 & 10.556 & 10.264 & $-0.007$ & $+0.001$ \\
 & & & & & \\
\hline
\end{tabular}
\end{center}

\pagebreak[4]
\begin{center}
TABLE 2 (con't)
\end{center}

\begin{center}
\begin{tabular}{lrrrrr}
\hline\hline
 & & & & & \\
Date & Star$^a$ & $J$ & $K'$ & $\Delta_J$ & $\Delta_K$ \\
 & & & & & \\
\hline
Aug 30 & 24 & 10.904 & 10.753 & $-0.011$ & $-0.011$ \\
 & & & & & \\
 & 23 & 12.997 & 12.374 & $-0.044$ & $-0.104$ \\
 & & & & & \\
 & 27 & 13.494 & 13.123 & $+0.021$ & $-0.058$ \\
 & & & & & \\
 & 30 & 11.923 & 12.015 & $+0.043$ & $+0.043$ \\
 & & & & & \\
 & 32 & 13.459 & 13.664 & $+0.051$ & $+0.047$ \\
 & & & & & \\
 & 7 & 11.105 & 10.940 & $-0.029$ & $+0.087$ \\
 & & & & & \\
Aug 31 & 29 & 13.175 & 13.346 & $-0.008$ & $+0.005$ \\
 & & & & & \\
 & 6 & 13.239 & 13.374 & $+0.001$ & $-0.046$ \\
 & & & & & \\
 & 2 & 10.713 & 10.466 & $+0.039$ & $+0.069$ \\
 & & & & & \\
 & 4 & 10.556 & 10.264 & $+0.042$ & $+0.017$ \\
 & & & & & \\
Sept 1 & 35 & 12.231 & 11.757 & $-0.039$ & $+0.003$ \\
 & & & & & \\
\hline
\end{tabular}
\end{center}

\noindent$^a$ FS number listed by Casali \& Hawarden (1992)

\pagebreak[4]
\begin{center}
TABLE 3

M13 NORMAL POINTS
\end{center}

\begin{center}
\begin{tabular}{ll}
\hline\hline
 & \\
$K$ & $J-K$ \\
 & \\
\hline
 & \\
10.0 & 0.75 $\pm 0.02$ \\
 & \\
10.5 & 0.72 $\pm 0.02$ \\
 & \\
11.0 & 0.68 $\pm 0.02$ \\
 & \\
11.5 & 0.62 $\pm 0.02$ \\
 & \\
12.0 & 0.59 $\pm 0.02$ \\
 & \\
12.5 & 0.57 $\pm 0.02$ \\
 & \\
13.0 & 0.55 $\pm 0.01$ \\
 & \\
13.5 & 0.52 $\pm 0.01$ \\
 & \\
14.0 & 0.50 $\pm 0.01$ \\
 & \\
14.5 & 0.47 $\pm 0.01$ \\
 & \\
\hline
\end{tabular}
\end{center}

\pagebreak[4]
\begin{center}
TABLE 3 (con't)
\end{center}

\begin{center}
\begin{tabular}{ll}
\hline\hline
 & \\
$K$ & $J-K$ \\
 & \\
\hline
 & \\
15.0 & 0.44 $\pm 0.01$ \\
 & \\
15.5 & 0.43 $\pm 0.02$ \\
 & \\
16.0 & 0.36 $\pm 0.02$ \\
 & \\
16.5 & 0.31 $\pm 0.02$ \\
 & \\
17.0 & 0.27 $\pm 0.02$ \\
 & \\
17.5 & 0.29 $\pm 0.02$ \\
 & \\
18.0 & 0.32 $\pm 0.02$ \\
 & \\
18.5 & 0.39 $\pm 0.02$ \\
 & \\
19.0 & 0.54 $\pm 0.07$ \\
 & \\
\hline
\end{tabular}
\end{center}

\pagebreak[4]
\begin{center}
TABLE 4

COLOR RELATIONS FOR SUB-DWARFS
\end{center}

\begin{center}
\begin{tabular}{rrr}
\hline\hline
 & & \\
$B-V$ & $J-K$ & $V-K$ \\
 & & \\
\hline
 & & \\
1.65 & 0.72 & 5.05 \\
 & & \\
1.55 & 0.70 & 4.40 \\
 & & \\
1.50 & 0.69 & 3.93 \\
 & & \\
1.35 & 0.68 & 3.35 \\
 & & \\
0.90 & 0.59 & 2.43 \\
 & & \\
0.80 & 0.56 & 2.10 \\
 & & \\
0.70 & 0.49 & 1.87 \\
 & & \\
0.60 & 0.43 & 1.63 \\
 & & \\
0.50 & 0.36 & 1.38 \\
 & & \\
0.40 & 0.28 & 1.23 \\
 & & \\
\hline
\end{tabular}
\end{center}

\pagebreak[4]
\begin{center}
TABLE 5

COLOR RELATIONS FOR GIANTS
\end{center}

\begin{center}
\begin{tabular}{rrr}
\hline\hline
 & & \\
$B-V$ & $J-K$ & $V-K$ \\
 & & \\
\hline
 & & \\
1.60 & 0.93 & 3.52 \\
 & & \\
1.50 & 0.87 & 3.34 \\
 & & \\
1.40 & 0.81 & 3.16 \\
 & & \\
1.30 & 0.77 & 3.04 \\
 & & \\
1.20 & 0.75 & 2.92 \\
 & & \\
1.10 & 0.70 & 2.77 \\
 & & \\
1.00 & 0.67 & 2.59 \\
 & & \\
0.90 & 0.63 & 2.43 \\
 & & \\
0.80 & 0.57 & 2.24 \\
 & & \\
0.70 & 0.48 & 1.99 \\
 & & \\
\hline
\end{tabular}
\end{center}

\pagebreak[4]
\begin{center}
TABLE 6

MASS FUNCTION EXPONENTS
\end{center}

\begin{center}
\begin{tabular}{rlc}
\hline\hline
 & & \\
Field & R$^a$ & $x_{ms}$ \\
 & & \\
\hline
 & & \\
2 & 3.9 & $+0.1 \pm 0.4$ \\
 & & \\
3 & 5.9 & $-6.1 \pm 5.0$ \\
 & & \\
4 & 3.4 & $-0.5 \pm 0.7$ \\
 & & \\
\hline
\end{tabular}
\end{center}

\noindent{$^a$} Distance from cluster center, in core radii, as presented in
Table 1.

\pagebreak[4]
\begin{center}
{\bf REFERENCES}
\end{center}
\parindent=0.0cm

Abia, C., \& Rebolo, R. 1989, ApJ, 347, 186

Alcaino, G., Liller, W., \& Alvarado, F. 1988, ApJ, 330, 569

Armandroff, T. E. 1989, AJ, 97, 375

Arp, H. C. 1955, AJ, 60, 317

Arp, H. C., \& Johnson, H. L. 1955, ApJ, 122, 171

Barbuy, B., \& Erdelyi-Mendes, M. 1989, A\&A, 214, 239

Baum, W. A. 1954, AJ, 59, 422

Baum, W. A., Hiltner, W. A., Johnson, H. L., \& Sandage, A. R. 1959, ApJ,
\linebreak[4]\hspace*{2.0cm}130, 749

Bell, R. A. 1992, MNRAS, 257, 423

Bell, R. A., \& Gustafsson, B. 1989, MNRAS, 236, 653

Bergbusch, P. A., \& VandenBerg, D. A. 1992, ApJS, 81, 163

Brown, J. A., Wallerstein, G., \& Oke, J. B. 1991, AJ, 101, 1693

Buckley, D. R., \& Longmore, A. J. 1992, MNRAS, 257, 731

Casali, M., \& Hawarden, T. 1992, JCMT-UKIRT Newsletter, 4, 33

Cohen, J. G. 1978, ApJ, 223, 487

Cohen, J. G., \& Frogel, J. A. 1982, ApJ, 255, L39

Cohen, J. G., Frogel, J. A., \& Persson, S. E. 1978, ApJ, 222, 165

Davidge, T. J., \& Simons, D. A. 1994a, ApJ, 423, 640

Davidge, T. J., \& Simons, D. A. 1994b, ApJ, 435, 207

Djorgovski, S., Piotto, G., \& Capaccioli, M. 1993, AJ, 105, 2148

Drukier, G. A., Fahlman, G. G., Richer, H. B., \& VandenBerg, D. A. 1988,
\linebreak[4]\hspace*{2.0cm}AJ, 95, 1415

Ferraro, F. R., Fusi Pecci, F., Montegriffo, P., Origlia, L., \& Testa, V.
1994,\linebreak[4]\hspace*{2.0cm}A\&A, in press

Folgheraiter, E. L., Penny, A. J., \& Griffiths, W. K. 1993, MNRAS, 264, 991

Frogel, J. A., Cohen, J. G., \& Persson, S. E. 1983, ApJ, 275, 773

Frogel, J. A., Persson, S. E., \& Cohen, J. G. 1983, ApJS, 53, 713

Guarnieri, M. D., Dixon, R. I., \& Longmore, A. J. 1991, PASP, 103, 675

Kadla, Z. I., Antal, M., Zdarsky, F., \& Spasova, N. 1976, Sov. Ast., 20, 403

Kraft, R. P., Sneden, C., Langer, G. E., \& Prosser, C. F. 1992, AJ, 106, 645

Kraft, R. P., Sneden, C., Langer, G. E., \& Shetrone, M. D. 1993, AJ, 106,
\linebreak[4]\hspace*{2.0cm}1490

Laird, J. B., Carney, B. W., \& Latham, D. W. 1988, AJ, 95, 1843

Langer, G. E., Suntzeff, N. B., \& Kraft, R. P. 1992, PASP, 104, 523

Larson, R. B. 1990, PASP, 102, 709

Lee, Y$-$W, Demarque, P., \& Zinn, R. 1990, ApJ, 350, 155

Leep, E. M., Wallerstein, G., \& Oke, J. B. 1986, AJ, 91, 1117

Leggett, S. K. 1992, ApJS, 82, 351

Lehnert, M. D., Bell, R. A., \& Cohen, J. G. 1991, ApJ, 367, 514

Lupton, R. H., \& Gunn, J. E. 1986, AJ, 91, 317

Pilachowski, C. A., Wallerstein, G., \& Leep, E. M. 1980, ApJ, 236, 508

Pryor, C., Smith, G. H., \& McClure, R. D. 1986, AJ, 92, 1358

Reid, N., \& Majewski, S. R. 1993, ApJ, 409, 635

Richer, H. B., \& Fahlman, G. G. 1986, ApJ, 304, 273

Richer, H. B., Fahlman, G. G., Buonanno, R., \& Fusi Pecci, F. 1990, ApJ,
\linebreak[4]\hspace*{2.0cm}359, L11

Rieke, G. H., \& Lebofsky, M. J. 1985, ApJ, 288, 618

Sandage, A. 1970, ApJ, 162, 841

Savedoff, M. P. 1956, AJ, 61, 254

Simons, D. A., Clark, C. C., Massey, S., Smith, S., \& Toomey, D. 1993, SPIE,
\linebreak[4]\hspace*{2.0cm}1946, 502

Stetson, P. B. 1987, PASP, 99, 191

Stetson, P. B., \& Harris, W. E. 1988, AJ, 96, 909

Straniero, O., \& Chieffi, A. 1991, ApJS, 76, 525

Suntzeff, N. B. 1981, ApJS, 47, 1

VandenBerg, D. A., \& Stetson, P. B. 1991, AJ, 102, 1043

VandenBerg, D. A., Bolte, M., \& Stetson, P. B. 1990, AJ, 100, 445

Wainscoat, R. J., \& Cowie, L. L. 1992, AJ, 103, 332

Wallerstein, G., Leep, E. M., \& Oke, J. B. 1987, AJ, 93, 1137

Webbink, R. F. 1985, in IAU Symp. 113, Dynamics of Star Clusters, ed.
\linebreak[4]\hspace*{2.0cm}J. Goodman, \& P. Hut (Dordrecht: Reidel), 541

Zinn, R., \& West, M. J. 1984, ApJS, 55, 45

\pagebreak[4]
\begin{center}
FIGURE CAPTIONS
\end{center}

FIG. 1.$-$Final $J$ images for Fields 1 (upper left hand corner), 2 (upper
right hand corner), 3 (lower left hand corner), and 4 (lower right hand
corner). In all cases North is to the left, and East is at the bottom. The
blank area in the Field 3 observations is due to a bad quadrant on the
detector.

\vspace{0.5cm}
FIG. 2.$-J$ (top panel) and $K$ (bottom panel) completeness curves for
Fields 1 (solid line) and 3 (dashed line).

\vspace{0.5cm}
FIG. 3.$-(K, J-K)$ CMDs for Fields 1 (upper left hand corner), 2 (upper right
hand corner), 3 (lower left hand corner), and 4 (lower right hand corner).
Only those stars with uncertainties in either magnitude of 0.07 magnitudes
or less, as derived by ALLSTAR, have been plotted.

\vspace{0.5cm}
FIG. 4.$-$Comparisons between the M13 normal point sequence
(solid curve) and other datasets. In the top panel the
observations made by Cohen et al. (1978) of bright M13 giants
are shown as open squares, while the ridgeline defined from photometry
of main sequence stars by Buckley \& Longmore (1992) is
shown as filled squares. In the lower panel, the
normal points for the cluster M4 by Davidge \& Simons (1994a) are shown as
solid squares. The M4 data have been shifted to match the color and brightness
of the M13 MSTO.

\vspace{0.5cm}
FIG. 5.$-$Comparison between M13 (crosses) and M4 (open squares) normal points,
based on observations obtained by Richer \& Fahlman (1986) and Alcaino et al.
(1988), respectively. The two brightest M13 points were obtained by averaging
the colors and brightnesses of the two SGB and three lower-RGB stars observed
by Richer \& Fahlman (1986). The M4 data were shifted to match the color and
brightness of the M13 MSTO.

\vspace{0.5cm}
FIG. 6.$-(B-V, J-K)$ relations for dwarfs (top panel) and giants (lower panel).
Note the different axial scales on the two panels.

\vspace{0.5cm}
FIG. 7.$-(B-V, V-K)$ relations for dwarfs (top panel) and giants (lower panel).
Note the different axial scales on the two panels.

\vspace{0.5cm}
FIG. 8.$-$Near-infrared isochrones derived from the sequences tabulated by
Straniero \& Chieffi (1991; top panel) and Bergbusch \& VandenBerg (1992; lower
panel). The sequences in the top panel are for 16 (solid line) and 20
Gyr (dashed line), while the lower panel shows sequences with ages of
12 (solid line) and 16 Gyr (dashed line). It is readliy evident that the
ability to detect a given age difference depends on the cluster age.

\vspace{0.5cm}
FIG. 9.$-$The M13 normal points (solid squares) compared with 12 Gyr (top
panel) and 16 Gyr (lower panel) [Fe/H] $\sim -1.6$ isochrones from Bergbusch \&
VandenBerg (1992). See text for additional details.

\vspace{0.5cm}
FIG. 10.$-$The M13 normal points (solid squares) compared with 16 Gyr (top
panel) and 20 Gyr (lower panel) [Fe/H] $\sim -1.6$ isochrones from Straniero
\& Chieffi (1991). See text for additional details.

\vspace{0.5cm}
FIG. 11.$-$16 and 20 Gyr near-infrared isochrones derived from the sequences
tabulated by Straniero \& Chieffi (1991), compared with data from Fields
2 ($K < 17.5$) and 3 ($K > 17.5$).

\vspace{0.5cm}
FIG. 12.$-$The completeness-corrected $K$ luminosity functions for Fields 2
(solid line), 3 (dashed line), and 4 (dashed-dotted line). The error bars
reflect counting statistics. $N_K$ is the number of stars per
square arcmin per 0.5 magnitude interval.

\vspace{0.5cm}
FIG. 13.$-$The mass functions for Fields 2 (solid line), 3
(dashed line), and 4 (dashed-dotted line). $N_M$ is the number of stars per
square arcmin per solar mass.
\end{document}